\documentclass[aps,prl,reprint,longbibliography,groupedaddress,floatfix]{revtex4-2}
\usepackage{amsmath,amssymb,graphicx,xcolor,hyperref,booktabs}
\DeclareFontShape{OMX}{cmex}{m}{n}{<-> cmex10}{}
\DeclareFontFamily{U}{msb}{}
\DeclareFontShape{U}{msb}{m}{n}{<-> msbm10}{}
\hypersetup{colorlinks=true,linkcolor=blue,citecolor=blue,urlcolor=blue}
\makeatletter
\AddToHook{env/thebibliography/begin}{\let\href\@secondoftwo\let\Eprint\@secondoftwo}
\makeatother
\graphicspath{{figures/}}
\newcommand{\PaperTitle}{Rydberg Constraint Ensembles for Nonnative Binary Optimization}
\begin{document}
% MAIN TEXT
\title{\PaperTitle}
\author{Zecheng Shen}
\email{oliver.shen@matriqcomputing.com}
\affiliation{MatriQ Computing, Binjiang District, Hangzhou, China}
\begin{abstract}
Rydberg blockade can exclude the configuration that minimizes a binary objective.
We expand its finite penalties into a weighted ensemble of exclusion graphs. The full ensemble preserves the ground energy, and the interactions remaining on each graph set its local detunings.
In simulations of two coupled squares, reversing a target coupling leaves the single-site energy slopes at uniform occupation unchanged, yet a collective mode of the residual interactions switches the favored diagonals from aligned to opposed.
On 32 held-out twelve-site targets, the simulated probability of recovering a ground state in 128 shots averages 69.82\% with residual detunings and 24.31\% with a tuned uniform pulse.
By scoring the pooled samples under the original objective, we can use these native layouts to search for ground states of a Hamiltonian that none implements alone.
\end{abstract}
\maketitle
\raggedbottom

\section*{Introduction}
Quantum computing aims to turn controlled quantum dynamics into capabilities beyond classical methods \cite{Preskill2018,Daley2022}.
A central algorithmic question is how to use native device operations without encoding the full computational task \cite{Abbas2024}.
In a programmable quantum simulator, the native interactions determine which spatial correlations develop \cite{Weimer2010Simulator,GeorgescuAshhabNori2014}.
A single image of a Rydberg array records a complete occupation pattern \cite{BrowaeysLahaye2020,Bernien2017,Ebadi2021Phases,Scholl2021Antiferromagnets}, which can be scored under an independently specified target energy \cite{McClean2021Mechanisms,Layden2023,WurtzSackWang2024}.
When no single layout implements the target Hamiltonian, how should an ensemble of layouts be driven so that its samples contain low-energy configurations of that target?

Many combinatorial optimization problems can be formulated as energy minimization over binary variables \cite{Lucas2014}.
Quadratic unconstrained binary optimization (QUBO) includes local fields and pair couplings \cite{GloverKochenbergerDu2019}, while its polynomial extension (PUBO) admits higher-order interactions \cite{BorosHammer2002,Nagies2025PUBO}.
A repulsive pair costs a finite energy, which a ground state can pay if occupying both sites gains more energy elsewhere.
Rydberg blockade suppresses simultaneous excitation of nearby atoms \cite{Jaksch2000Blockade,Lukin2001Blockade,Urban2009Blockade,Gaetan2009Blockade,SaffmanWalkerMolmer2010}. A single ideal blockade graph can therefore exclude the configuration of lowest target energy \cite{pichler,Pichler2018Complexity,ebadi}.
Quantum wires and auxiliary constraints extend the class of objectives that can be encoded directly \cite{Qiu2020Wires,Kim2022Wires,Oliveira2026SparseWires,nguyen,Byun2024HigherOrder,Byun2024QUBO,Gloeckner2026Constraints}.
Parity architectures encode distant couplings in local constraints \cite{LechnerHaukeZoller2015,Glaetzle2017Annealer,Lanthaler2023}. Hybrid algorithms can also use native samples as information about a nonnative problem \cite{WurtzSackWang2024}.
Rearranging the atoms changes which occupation pairs are blocked \cite{Nogrette2014Geometry,Endres2016Assembly,Barredo2016Assembly,Henriet2020}, so different layouts can admit different configurations of the same objective.

A layout can admit a ground state yet rarely produce it. In Rydberg independent-set studies, the probability of finding an optimum depends on the annealing schedule \cite{ebadi,Kim2024King}. Local detunings can encode vertex weights \cite{oliveira} or favor occupations suggested by graph connectivity \cite{yeo}. With competing couplings, the cost of a site excitation depends on the occupations its layout permits.

Here we express selected finite penalties as a weighted ensemble of exclusion graphs, drawing on auxiliary-bond representations in statistical mechanics \cite{FortuinKasteleyn1972,es}. The remaining terms define a residual energy. For enumerable families, we visit graphs in decreasing posterior weight and retain the legal strings of lowest residual energy on each. In Rydberg simulations, the residual on each layout sets its detunings. On two coupled squares, a collective mode of the residual reverses the relative polarization when a target coupling changes sign, even though the single-site slopes at uniform occupation are unchanged. The driven layouts thus supply candidates for an objective that none implements in full, and the original energy identifies the best string in their pooled samples.

\begin{figure}[tb]
\centering\includegraphics[width=\columnwidth]{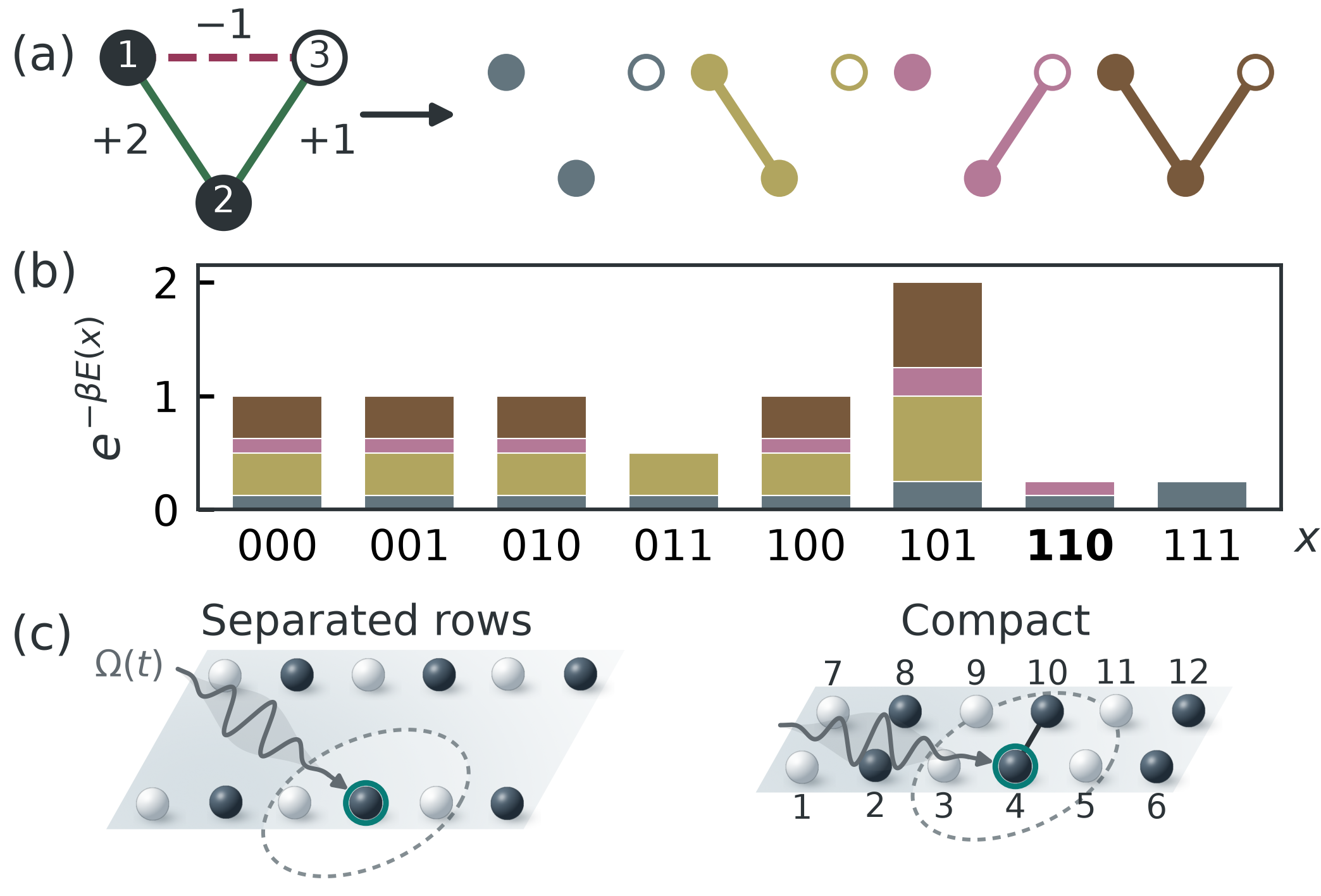}
\caption{\textbf{From exclusion graphs to a Rydberg drive.} (a) A signed triangle and its four exclusion graphs for \(x=110\). (b) Their contributions to \(e^{-\beta E(x)}\) at \(\beta=\ln 2\), colored as in (a). (c) Both layouts share the drive \(\Omega(t)\) and show the same target ground state (dark atoms). Site numbers in the compact layout apply to both arrays. Only in the compact layout does site 10 lie inside the blockade disk around site 4.}\label{fig:access}
\end{figure}

\section*{Weighted graph ensembles}
The triangle in Fig.~\ref{fig:access}(a) has energy \(E(x)=2x_1x_2+x_2x_3-x_1x_3\) for binary occupations \(x_i\in\{0,1\}\). We can impose or omit an exclusion on each repulsive bond. All four resulting graphs retain the same residual attraction \(A(x)=-x_1x_3\).
For a family \(F\) of selected site sets with nonnegative penalties \(w_e\), let \(m_e(x)=\prod_{i\in e}x_i\) and \(A(x)=E(x)-\sum_{e\in F}w_em_e(x)\).
At inverse ranking temperature \(\beta>0\), impose exclusion \(e\) with probability \(p_e=1-e^{-\beta w_e}\). The single-penalty identity is
\begin{equation}
e^{-\beta w_e m_e(x)}=(1-p_e)+p_e[1-m_e(x)] .
\end{equation}
A fully occupied set retains weight only when the exclusion is absent. Expanding independently over sets gives a graph prior \(P_0(G)\) and a legality indicator \(I_G(x)\), with
\begin{equation}
e^{-\beta E(x)}=\sum_{G\subseteq F}P_0(G)e^{-\beta A(x)}I_G(x).
\label{eq:factor}
\end{equation}
This construction belongs to a family of graphical expansions relating interaction weights to discrete constraints \cite{FortuinKasteleyn1972,es,mak,MakSharma2007,Sasaki2010}.
The ground state \(101\) has the largest Gibbs weight, with \(E=-1\). The highlighted \(110\) pays the repulsion on bond \((1,2)\), giving \(E=+2\). At \(\beta=\ln 2\) the two graphs admitting it each contribute \(1/8\) to its weight [see Fig.~\ref{fig:access}(b)].
Summing over strings gives the graph posterior \(P_\star(G)\propto P_0(G)\sum_x e^{-\beta A(x)} I_G(x)\). Sampling this posterior and then its conditional legal Boltzmann law reproduces the target distribution \cite[see Supp.~S1]{PaperISupplement}.

The same decomposition preserves the original ground energy \(E^*=\min_xE(x)\),
\begin{equation}
 E^*=\min_{G\subseteq F}\left[\min_{x:I_G(x)=1}A(x)+\sum_{e\in F\setminus G}w_e\right].
 \label{eq:bridge}
\end{equation}
Only omitted penalties can be paid by a legal string. For an original ground state, imposing every exclusion it obeys makes the bound from those omitted penalties tight. One exact legal residual minimizer per graph therefore suffices across the complete family [see Supp.~S2].
An exact optimization over the complete graph family recovered a certified ground state of a 128-variable rewired ring [see Supp.~S3].
To truncate this search, we visit graphs in decreasing posterior weight, retain up to \(K\) legal strings of lowest residual energy per graph, and score their union under \(E\). We test this ordering on twelve disordered eight-site targets. For binary occupations \(x_i\), each quadratic target has energy
\begin{equation}
 E(x)=\sum_{i=1}^{N}h_i x_i+\sum_{i<j}J_{ij}x_ix_j ,
 \label{eq:quadratic}
\end{equation}
where \(h_i\) is the site energy and \(J_{ij}\) the net pair coupling. For \(N=8\) we also test each target with a cubic term \(\pm4x_1x_5x_8\). At unit ranking temperature and \(K=1\), the first two of 1,024 ranked graphs suffice to recover a ground state for every quadratic target and every cubic extension. This enumeration uses offline partition sums and exact residual minimization on each visited graph. Certifying the minimum requires a bound on the unvisited graphs [see Supp.~S3].

\section*{Graph-conditioned detunings}
The posterior orders graphs at equilibrium. To use one of them as a source of measured strings, we need a drive for the energy that remains on its legal configurations. Starting with \(A\) from Eq.~\eqref{eq:factor}, we remove every monomial containing an exclusion of \(G\). The resulting \(A_G\) is the unique expansion in the surviving monomials that agrees with the residual on every legal occupation. If reassigning variables to sites adds a penalty to the graph construction, we subtract the same term from \(A\) and undo the permutation before scoring. Every layout then refers to the same \(E\).
At uniform reference occupation \(\rho\), the derivative \(g_i^G=(\partial A_G/\partial x_i)_{\rho\mathbf{1}}\) gives the mean residual cost of exciting site \(i\) when other sites are occupied independently.

Figure~\ref{fig:access}(c) shows how geometry changes the residual for a fixed twelve-site quadratic target. Its pair coefficient \(J_{4,10}=+1\) combines a selected penalty \(+4\) and a residual attraction \(-3\). Increasing the row spacing moves site 10 outside the blockade radius of site 4, allowing both sites to be occupied. The compact graph excludes every ground state, while separating the rows admits two. The attraction vanishes on every legal string of the compact graph. Opening the rows restores its contribution of \(-3/4\) to each endpoint slope at \(\rho=1/4\). In our pulse, lower residual cost raises the final detuning \cite{oliveira,yeo}. Site 4 thus shifts from \(3.1875\) to \(3.7500\) MHz on the same objective. An independent eight-site example shows the corresponding change in geometric access [see Fig.~S2].
Coherent drives can change the ordering dynamics and revivals in blockade arrays \cite{Keesling2019Dynamics,Bluvstein2021Control}. With global Rabi drive \(\Omega(t)\), occupation \(n_i\), spin flip \(\sigma_i^x\) and native repulsions \(V_{ij}\), the Hamiltonian reads \cite{Labuhn2016Ising,Lienhard2018Correlations,DeLeseleuc2018Mapping}
\begin{equation}
 H(t)/\hbar=\sum_i\left[\Omega(t)\sigma_i^x/2-\Delta_i(t)n_i\right]
 +\sum_{i<j}V_{ij}n_in_j.
\end{equation}
A fixed spatial mask and a time-dependent amplitude set the local part of \(\Delta_i(t)\). Starting from the empty array, a pulse \(\phi\) on layout \(G\) produces a nonequilibrium law \(q_{G,\phi}(x)\) with the full \(r^{-6}\) repulsion between every pair. Candidates from every layout are compared under the same \(E(x)\), including strings outside the layout's ideal legal subspace when raw readout is used.

\section*{Control of collective polarization}
\begin{figure}[tb]
\centering\includegraphics[width=\columnwidth]{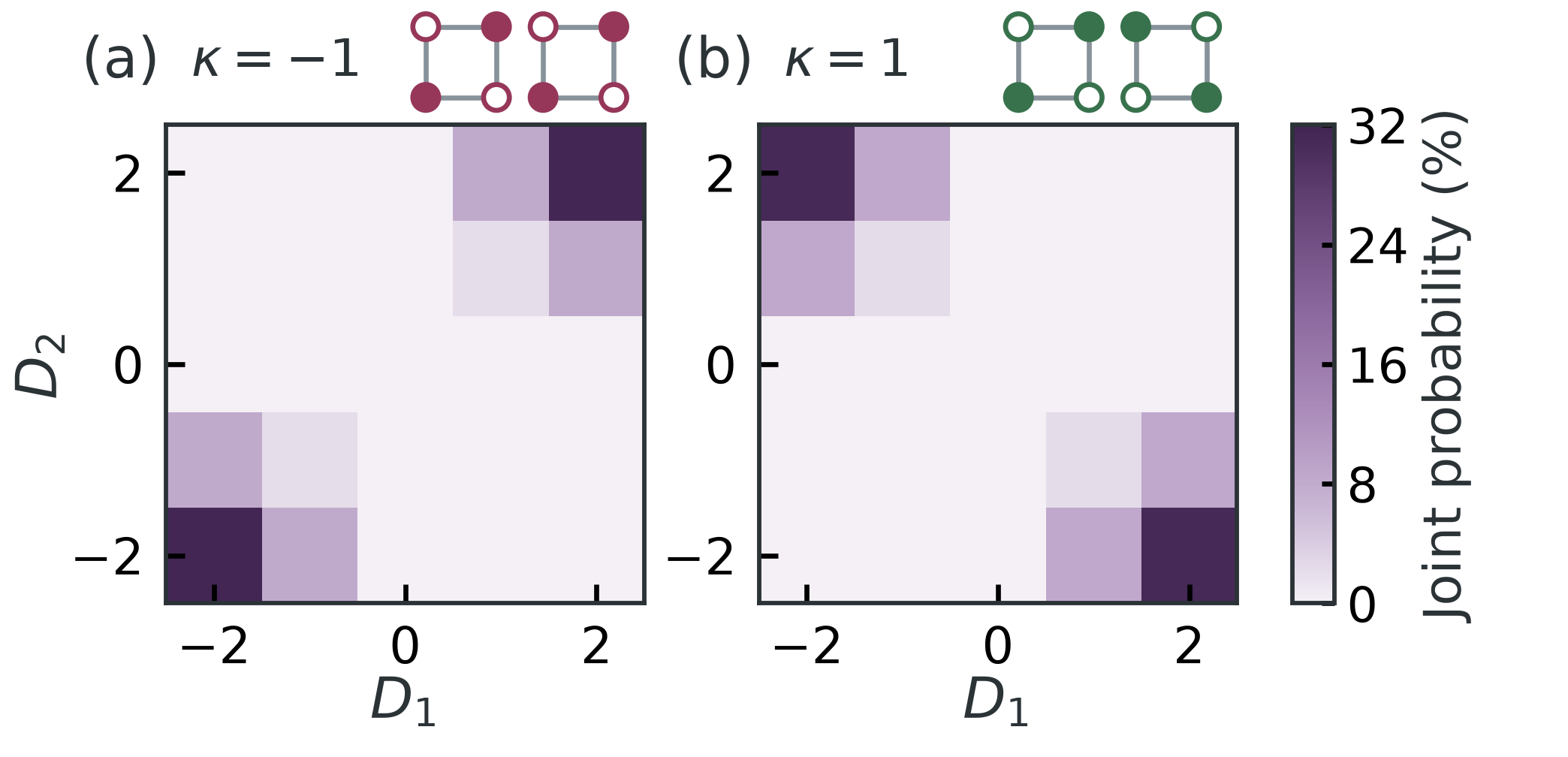}
\caption{\textbf{Reversing a target coupling reverses the relative polarization.} Joint distributions of the diagonal imbalances \(D_1,D_2\) for (a) \(\kappa=-1\) and (b) \(\kappa=+1\). Filled sites above each map show one ground state. Both maps use a common probability scale and average equally over two symmetry-related masks at \(20\,\mu{\rm m}\) separation and \(3.95\,\mu{\rm s}\) pulse duration.}\label{fig:mode}
\end{figure}

The twelve-atom example shows how a local detuning follows the energy cost of a single excitation. A pair energy can also select which diagonal of a blockade square fills, even when the site slopes favor the other diagonal. To isolate this collective choice, we arrange eight atoms in two coupled squares. We can then reverse the preferred relative orientation of their diagonals while keeping the geometry and all single-site slopes at uniform occupation fixed. Numbering the sites around each square \(b\), define its diagonal imbalance \(D_b=x_{b1}-x_{b2}+x_{b3}-x_{b4}\). Filling one diagonal gives \(D_b=+2\) and filling the other gives \(-2\). We score each configuration with the target energy
\begin{equation}
 E_\kappa(x)=-\sum_i x_i+4\sum_{\langle ij\rangle}x_ix_j
             +\frac{\kappa}{4}D_1D_2 ,
 \label{eq:square-target}
\end{equation}
where \(\langle ij\rangle\) runs over the eight square sides. Filling one diagonal on each square gains four units of energy without paying a side penalty. For these configurations \(D_1D_2/4\) is \(+1\) for aligned diagonals and \(-1\) for opposed diagonals. The target thus favors alignment for \(\kappa=-1\) and opposition for \(\kappa=+1\) at the same native pair interactions [see Fig.~\ref{fig:mode}]. The intersquare contribution to the site slope is \(\kappa\eta_i D_{\rm other}/4\), with \(\eta_i=+1,-1,+1,-1\) around each square. At uniform occupation, \(D_{\rm other}=0\), so reversing \(\kappa\) leaves the site slopes unchanged.

Each target has two ground states of energy \(-5\), and the two sets are disjoint. For a single output law that is blind to \(\kappa\), let \(p_-\) and \(p_+\) be the probabilities assigned to these sets. Their sum cannot exceed one. With equally likely signs, the single-shot bound is
\begin{equation}
 (p_-+p_+)/2\leq 1/2. \label{eq:signbound}
\end{equation}

In the two normalized alternating directions, the projected residual coupling matrix has zero diagonal and off-diagonal element \(\kappa\).
Its lowest mode selects the relative sign of the two alternating detunings. Reversing the mode gives the symmetry-related mask for the same target. We sample both masks with equal probability at \(20\,\mu{\rm m}\) separation and \(3.95\,\mu{\rm s}\) pulse duration. For either sign of \(\kappa\), this mixture gives \(\langle D_1\rangle=\langle D_2\rangle=0\). Reversing the coupling shifts the weight in Fig.~\ref{fig:mode} from aligned to opposed diagonals. Because the two means vanish, the normalized covariance equals the mean product \(\langle D_1D_2\rangle/4\), which changes from \(+0.804\) to \(-0.798\).

How much of this covariance remains when the mask is held fixed? Let \(\langle D_b\rangle_m\) be the mean imbalance on square \(b\) for mask \(m\). At each target sign, the total covariance separates as
\begin{equation}
\begin{split}
{\rm Cov}(D_1,D_2)={}&{\mathbb E}_m[{\rm Cov}(D_1,D_2\mid m)]\\
 &+{\rm Cov}_m(\langle D_1\rangle_m,\langle D_2\rangle_m).
\end{split}
\end{equation}
For \(\kappa=-1,+1\), the within-mask term divided by four is \(-0.00032\) and \(+0.00052\), respectively. The corresponding between-mask contributions, \(+0.80449\) and \(-0.79847\), account for nearly the entire reversal.

If the signal comes from the polarizations written by the masks, it should survive independent sampling of the sites within each mask. For each mask \(m\), we form the product distribution \(q_m(x)=\prod_i p_{im}^{x_i}(1-p_{im})^{1-x_i}\) from the native site occupations \(p_{im}=\langle x_i\rangle_m\). Mixing these two distributions reproduces the reversal and gives 64.14\% mean single-shot ground-state recovery across the two signs. At the same separation the native masks give 62.32\%, compared with 49.89\% for the strongest uniform pulse in the scan. A hard-core law fitted to the same marginals reproduces the independent-source success rates to two decimal places [see Supp.~S4]. The target mode selects the relative polarization. How readily the pulse prepares it depends on the geometry, with native recovery falling to 37.72\% at \(16\,\mu{\rm m}\).

\section*{Low-energy sampling and readout}
\begin{figure}[tb]
\centering\includegraphics[width=\columnwidth]{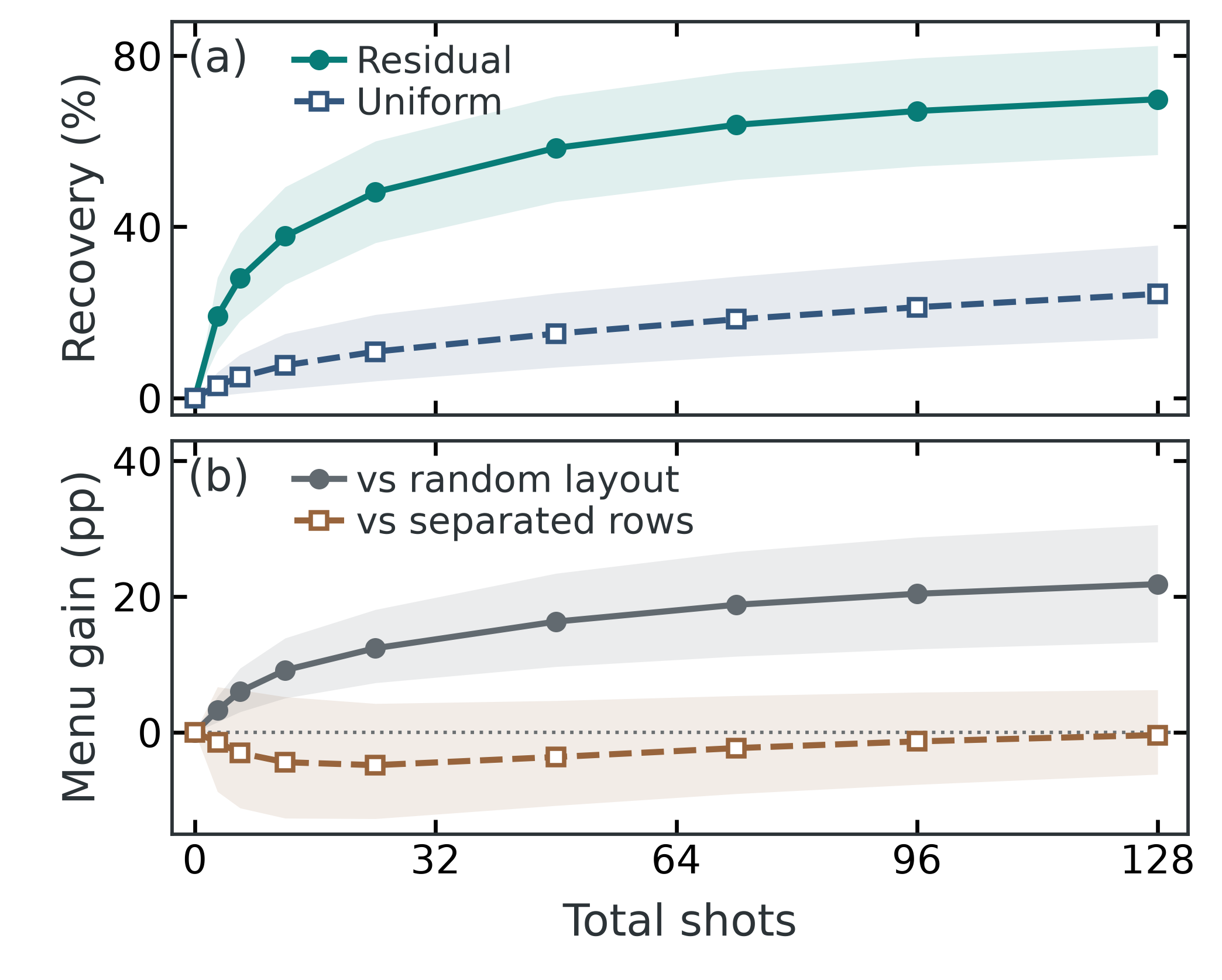}
\caption{\textbf{Finite-shot ground-state recovery on twelve-atom arrays.} Both panels average over the same 32 held-out targets. (a) Recovery for residual and uniform fields, dividing shots as evenly as possible among three layouts. All distinct strings are scored. (b) Recovery of the residual three-layout menu minus that of a single layout, chosen uniformly once per run or fixed to separated rows. Points give the evaluated budgets. Bands are pointwise 95\% intervals from 10,000 target resamples stratified by family, paired in (b).}\label{fig:recovery}
\end{figure}

The eight-atom model isolates how a residual coupling selects the relative orientation of two squares. To test whether residual detunings improve the probability of sampling a ground state with a finite number of shots, we return to the disordered twelve-atom arrays of Fig.~\ref{fig:access}(c). Here disorder makes the residual slopes site dependent. We compare the resulting fields with a uniform pulse on 32 held-out quadratic targets of the form in Eq.~\eqref{eq:quadratic}, sixteen from each of the attractive and frustrated families. Eight separate development targets select the uniform endpoint, 8 MHz, at the upper end of the scan. Both fields use \(3\,\mu{\rm s}\) pulses on the same three layouts. With 128 shots divided as 43/43/42 across the layouts, we score every distinct observed string under the original energy, so any measured ground state is retained. Under this readout the residual fields raise recovery from 24.31\% to 69.82\%, a paired gain of 45.51 percentage points (95\% interval, 30.40 to 59.98) [see Fig.~\ref{fig:recovery}(a)]. The uniform and residual fields require comparable numbers of energy evaluations, averaging 35.11 and 34.50, respectively. Flattening each residual field to its spatial mean lowers recovery to 16.54\%, even as the mean number of strings scored rises to 79.74.

The uniform comparison leaves open which energy should determine the spatial variation. We compare fields derived from the full objective and its residual at fixed layouts, variable assignment and 128-shot allocation. Both fields use the same quarter-occupation slope rule and omit monomials forbidden on the given graph. The full-objective field retains the selected penalties on pairs left open by that graph. The residual field removes those penalties before taking the slope. For this comparison we score the three most frequent legal strings per layout under the original objective. This common readout gives 31.45\% ground-state recovery with the full-objective field and 40.65\% with the graph-conditioned residual field, a paired gain of 9.2 percentage points on the 32 targets (95\% interval, 2.6 to 17.0) [see Supp.~S5].

Once the residual fields are fixed, the three layouts differ in the probability they assign to a ground state. Scoring every distinct observed string isolates the effect of allocating shots among the layouts [see Fig.~\ref{fig:recovery}(b)]. At 128 shots the three-layout menu gives 69.82\% recovery, compared with 47.99\% for a uniformly chosen single layout (paired gain 21.83 points, 95\% interval 13.30 to 30.58). These policies use 34.50 and 25.71 mean energy evaluations. Separated rows give the largest mean among the three tested fixed layouts, 70.23\% with 33.61 calls. The three-layout menu differs from the separated rows by \(-0.41\) points in recovery (95\% interval, \(-6.20\) to \(6.25\)). Opening the other layouts gives 28 of the 32 targets access to a ground state, compared with 26 for the separated rows. On a 32-site chorded cycle, simulated readout with residual detunings recovered both certified ground states [see Supp.~S5].

The native sampler takes 128 shots per target and evaluates the original energy only for distinct strings on the shortlist. With a budget of 128 original-energy evaluations per target, annealing and tabu search recover a ground state in 72.85\% and 80.98\% of runs, respectively. To isolate the role of correlations within a layout, we compare sampling distributions on a separate cohort at matched scoring cost. Native samples give 6.2 percentage points higher recovery than independent bits with the same site occupations. A hard-core model fitted offline to the exact one-site marginals enforces the layout exclusions and exceeds native recovery by 3.4 points.

At readout, the per-layout shortlist ranks measured strings by frequency. The exact graph search ranked legal strings by residual energy. Even a ground state present on every layout can be discarded by the frequency cutoff. On one frustrated twelve-site target an energy \(-7\) optimum appeared two, three and two times, yet no layout retained it among its three most frequent strings.
On sixteen frustrated twelve-site targets, we compare list depths using identical samples [see Supp.~S6]. Retaining three, ten or all distinct strings per layout gives 57.57\%, 75.68\% and 86.77\% recovery at mean costs of 6.8, 21.3 and 33.7 energy evaluations.

Deepening the list recovers a string already measured. Sampling a different source can reveal a previously unseen string. On 160 four-site targets, for example, a fifth shot from the residual slope after four strong uniform shots adds 8.58 percentage points of recovery, compared with 2.67 for repeating the pulse. At sixteen shots two global pulses already recover 96.36\%, comparable to 95.24\% for a three-source mixture [see Supp.~S7]. These estimates assume independent shots. In the four-site noise simulation, recovery falls when the same control error persists across shots [see Supp.~S8].

\section*{Conclusion}
A blockade array can search a nonnative energy landscape by separating three choices. The constraint ensemble identifies layouts that admit relevant configurations. Within an accessible sector, the residual gives the drive a spatial direction, including a relative polarization invisible to local slopes. Each layout's ground-state weight sets the expected return from the shots assigned to it. The scoring budget sets how many measured strings can be tested for a rare optimum. The same target objective can thus guide both preparation and selection \mbox{across physical layouts}.

\section*{Data and Code Availability}
The data and code supporting the findings of this study are available from the author upon reasonable request.

% BibTeX formatting control; this is not a numbered reference.
\makeatletter
\immediate\write\@auxout{\string\citation{PaperIStyle}}
\makeatother
% Works cited only in the Supplemental Material.
\nocite{Byun2022Platonic,Jeong2023SAT,StastnyBuchlerLang2023,Pan2026Triangular}
\nocite{Schuetz2025Compilation,Cazals2025Gadgets,Dlaska2022ParityGates,BloqadeCaps}
\nocite{AquilaLD,Schauss2012Order,Schauss2015Crystallization,Guardado2018Quench}
\nocite{Omran2019Cat,Perseguers2025,DeLeseleuc2018Imperfections,Steinert2023Interactions}
\nocite{jung2026,Serret2020,Ronnow2014,Proctor2025Benchmark}
\nocite{metropolis,hastings,GemanGeman1984,Schuetz2026QreduMIS}
\nocite{VeachGuibas1995,OwenZhou2000,Cappe2008,heowen}
\nocite{BubeckErnstGarivier2013,Aquila2023}
\bibliography{references}
\end{document}